\documentclass[12pt,epsf]{article}

\usepackage{latexsym}
\usepackage{epsfig}
\usepackage{amsfonts}

\begin{document}

\begin{center}
{\bfseries  \Large Casimir energy in the compact QED on the lattice}
\vskip 5mm

Oleg Pavlovsky$^{\dag}$ and Maxim  Ulybyshev$^{\ddag}$

\vskip 5mm

{\small {\it Institute for Theoretical Problems of Microphysics,
Moscow State University, Moscow, Russia }
\\
$\dag$ {\it E-mail: ovp@goa.bog.msu.ru } $\ddag$ {\it E-mail:
ulybyshev@goa.bog.msu.ru }}
\end{center}

\vskip 5mm

\begin{center}
\begin{minipage}{150mm}
\centerline{\bf Abstract} A new method based on the Monte-Carlo
calculation on the lattice is proposed to study the Casimir effect
in the compact lattice $U(1)$ theory with Wilson action. We have
studied the standard Casimir problem with two parallel plane
surfaces (mirrors) and oblique boundary conditions on those as a
test of our method. Physically, this boundary conditions may appear
in the problem of modelling of the thin material films interaction
and are generated by additional Chern-Simons boundary term. This
approach for the boundary condition generation is very suitable for
the lattice formulation of the Casimir problem due to gauge
invariance. This method can be simply generalized on the case of
more complicated geometries and other gauge groups.
\end{minipage}
\end{center}

\vskip 3mm


\vskip 3mm


\vskip 10mm

%%%%%%%%%%%%%%%%%%%%%%%%%%%%%%%%%%%%%%%%%%%%%%%%%%%%%%%%%%%%%%%%%%%%%%%%%%%%%%%
\section{Introduction and general motivation}
During the last few years Casimir effect has attracted much
attention due to the great experimental and theoretical progress in
studying of this phenomenon. This macroscopic quantum effect  plays
crucial role in nanophysics, micro-mechanics, quantum optics,
condensed matter physics, material science, and also it is very
important for different models of boundary states in hadron physics,
heavy-ion collisions, and cosmology.

Nowadays, many theoretical methods for calculation of Casimir effect
are proposed. Unfortunately, most of these methods are based on the
fixed boundary conditions or external potentials. Such a
simplification can lead to problems in gauge invariance,
renormalizability and locality. Moreover, there are some problems
with different thermal corrections to Casimir force: different
methods of calculations of the Casimir effect predict different
corrections. Another part of the problem is calculation of Casimir
force for complicated forms of boundary surfaces. Typically, various
approximate methods (like the proximity force approximation method
\cite{bordag-mastepanenko}, \cite{Deryagin}) are used in the case of
curved surfaces, but it is still unclear now whether those
approximations are correct and for what tasks they can be applied.
And finally, analytic methods for Casimir effect calculation are
very complicated and strongly dependent on the shape of surfaces.
Practically, Casimir effect was studied analytically only for cases
of plane, spherical and cylindrical surface forms. More complicated
tasks have not been studied very well by this moment but such cases
typically appear in experiments.

Another very interesting and important problem is the calculation of
Casimir effect for other gauge groups (for example, SU(2) and SU(3))
and for fermion fields; both are very important for particle physics
and for cosmology.

Based on arguments discussed above, it seems a very important task
now to create a general method for calculation of Casimir effects
which would work well for different shapes of boundary surfaces, for
different fields, at non-zero temperature and density and under
other external factors. It means that such a method should be
formulated very generally for working in different coupling regimes
and different external conditions. And we believe that only direct
lattice calculations in quantum field theory can meet all these
requirements.

In our paper we consider the simplest Casimir problem with two
parallel planes as a test of our method. The existence of the
analytical answer for this problem is the additional motivation for
such a choice of bound surfaces.  This analytical  answer assists us
in fitting procedure for our numerical results.

 A crucial obstacle on the way of the realization of any Casimir problem on
the lattice is the following. If the Casimir energy is a reaction of
the vacuum on the presence of the boundary, what is "the boundary"
in terms of lattice formalism? In other words, what is an observable
quantity corresponding to such boundary? In fact, the answer is
non-trivial and we devote the first part of the paper to this
question. In the second part, we discuss the lattice algorithms and
numerical results. And the last part is a conclusion.

\section{Chern-Simons boundary conditions and Casimir effect}

Casimir effect is a reaction of the vacuum on boundary condition.
The spectrum of vacuum fluctuations depends on the boundary
conditions. Changing of the boundary conditions leads to changing of
the spectrum of vacuum fluctuations and so to generating of the
corresponding Casimir force on the boundary. In the standard quantum
field theory formalism, such changing of spectrum of vacuum
fluctuations can be described, for example, by means of Green
function method \cite{bordag-mastepanenko}. This approach is a very
powerful tool for studying many essential Casimir tasks
\cite{bordag-mastepanenko}. Unfortunately, the application of this
analytical method to the case of more complicated shape of the
boundary surfaces is not so easy due to calculation difficulties.
Our aim is the creation of the numerical method for the Casimir
effect calculation directly from the quantum field theory action.
The lattice formalism looks very attractive for this role but
manifestly we can not base in our approach on the separation of
vacuum modes corresponding with boundary from the full spectrum of
vacuum fluctuation. In lattice formalism we work in Euclidian space
and deal with full spectrum of vacuum fluctuations and can not
easily snatch out vacuum fluctuations corresponding to some boundary
conditions. We need some very delicate approach for separation of
vacuum fluctuation modes that preserve gauge invariance of our
lattice formalism. Fortunately, such an approach to Casimir problem
was proposed recently \cite{bordag-vasilevich, vasilevich}.

This approach is based on a very elegant idea coming from some
unique properties of the Chern-Simons action in three dimensions
\cite{bordag-vasilevich, vasilevich}. Let us consider
electro-magnetic fields in 3+1 dimentions with the Maxwell action
and additional Chern-Simons action given on 3-dimentional integral
on the boundary surface $S$:
\begin{equation}
S=-\frac{1}{4} \int d^4x\  F_{\mu\nu}F^{\mu\nu} -
 \frac{\lambda}{2} \oint d^3s \,
    \varepsilon^{\sigma \mu \nu \rho}
n_{\sigma } A_{\mu} ( x ) F_{\nu \rho} ( x )  , \label{action}
\end{equation}
where $\varepsilon^{\sigma \mu \nu \rho}$ is the Levi-Civita tensor
and $n_{\sigma }$ is  the normal vector to the boundary surface $S$,
$\lambda$ is a real parameter.

Let us consider now the simplest form of boundary surface $S$,
namely two parallel infinite planes placed at the distance $R$ from
each other. The Chern-Simons formulation of this canonical Casimir
problem was studied analytically in series of works \cite{markov1,
markov2}. We will use this analytical answer for the fitting of our
numerical data.

In the case of plane form of the boundary surface $S$ the
Chern-Simons action in (\ref{action}) has the following form:
\begin{equation}
 S_{CS}=\frac{\lambda}{2}\int(\delta(x_3)-\delta(x_3-{R}))
 \varepsilon^{3 \mu\nu \rho} A_{\mu}(x) F_{\nu
\rho}(x)d^4x. \label{actionCS}
\end{equation}
where, in our formulation of this Casimir problem, normal vectors to
the planes are turned in the opposite directions. This choice of the
normal vector orientation corresponds to our renormalization
procedure based on the connection between open and closed Casimir
problems.

If the parameter $\lambda$ is small, electro-magnetic fields
obviously don't feel any boundary and are free. What happens if the
parameter $\lambda$ becomes large and tends to infinity and  fields
dynamics on the boundary surface $S$ is determined by Chern-Simons
action? Let us consider the equation of motion obtained from the
action (\ref{action}):
\begin{eqnarray}
\Box A^\mu + \lambda (\delta(x_3)-\delta(x_3-{R}) )\varepsilon^{ 3
\sigma  \nu \rho} A_\sigma
\partial_\nu A_\rho  =0 . \label{equa_mot}
\end{eqnarray}
At $\lambda \rightarrow \infty$, it is easy to obtain from
(\ref{equa_mot}) a corresponding boundary conditions on the surface
$S$:
\begin{eqnarray}
&&E_n\vert_{S}=0 ,\qquad H_\parallel \vert_{S}=0 ,\label{condition}
\end{eqnarray}
where $E_n$ and $H_\parallel$ are normal and longitudinal components
of electric and magnetic fields correspondingly. These conditions
mean the nulling of the energy flux of the electromagnetic field
through the surface.
\begin{figure}[t]
\begin{center}
 \epsfbox{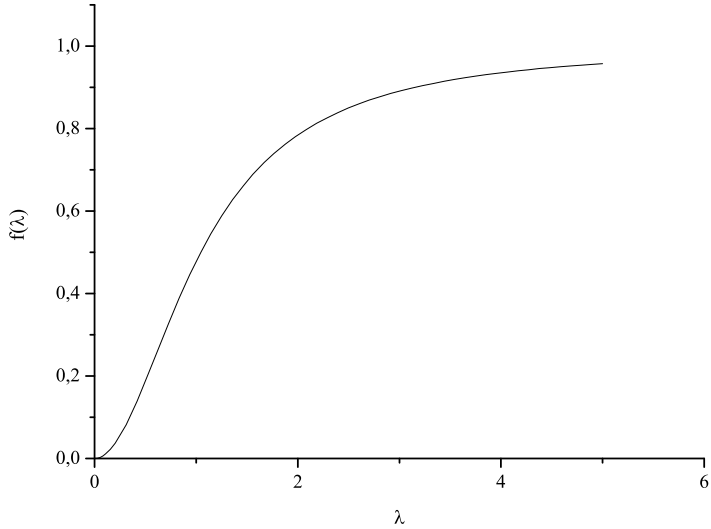} \end{center} \caption{ $f(\lambda)$. }
\end{figure}

This phenomenon corresponds to the well-known property of the
Chern-Simons theory in 3-dimension, namely to topological vortex
(strings) generation \cite{vortex, vortex2}. As it was shown by
Witten \cite{witten}, Chern-Simons theory, which is a quantum field
theory in three dimensions, is exactly solvable by using
nonperturbative methods and the topological vortices play a leading
role in this process. It was shown that the partition function of
this theory depends on the topology of the surface $S$ and gauge
group only, and tends to the zero if $\lambda$ tends to infinity.
There are no propagation modes in this case and if the boundary
surface $S$ is closed, the gauge fields inside and outside this
surface become separated from each other by the topological vortex
fields on the surface. In the case of the finite value of $\lambda$,
we have a non-trivial interaction between topological vortices and
electro-magnetic waves \cite{vortex2}. For our practical purposes it
is enough to know that the parameter $\lambda$ plays the role of the
effective regulator of the penetrability of our boundary surface for
electro-magnetic wave.

The analytical answer of Casimir energy per unit area for two planes
is given by Green function method is the following \cite{markov1,
markov2}:
\begin{eqnarray}
 E_{Cas}=-\frac{\pi^2}{720 R^3} f(\lambda), \label{fullCas}
\end{eqnarray}
where function $f(\lambda): \lim_{\lambda \rightarrow \infty}
f(\lambda) = 1$ (Fig. 1) can be written as:
\begin{eqnarray}
 f(\lambda)=\frac{90}{\pi^4}\mbox{Li}_4\left(\frac{\lambda^2}{\lambda^2+1}\right). \nonumber
\end{eqnarray}
here the polylogarithm function $\mbox{Li}_4(x)$  is defined as
$$
\mbox{Li}_4(x)=\sum_{k=1}^\infty \frac{x^k}{k^4}=-\frac{1}{2}\int_0
^\infty k^2 \ln(1-xe^{-k})dk.
$$

In our paper we will study the Casimir energy per unit area behavior
at small $\lambda$, which is the following:
\begin{eqnarray}
 E_{Cas}=-\frac{\lambda^2}{8 \pi^2 R^3}+ O(\lambda^4)
 \label{assimpt}
\end{eqnarray}

%%%%%%%%%%%%%%%%%%%%%%%%%%%%%%%%%%%%%%%%%%%%%%%%%%%%%%%%%%%%%%%%%%%%%%%%%%%%%%%
\section{Wilson "bag" and numerical lattice simulation of Maxwell-Chern-Simons theory}
In our paper we use the four-dimensional hyper-cubical lattice and
the simplest form of the action for the U(1) gauge fields (so-called
"Wilson action"):

$$
S_W=\beta \sum_x \sum_{\mu \nu} (1-\mbox{Re\,} U_{p,x,\mu\nu}),
$$
where the link and plaquette variables are defined as:
$$U_{l,x,\mu}=e^{i g A_{\mu}(x) a},$$
$$U_{p,x,\mu\nu}=U_{l,x,\mu} \, U_{l,x+\hat{\mu},\nu} \,
U^{\dagger}_{l,x+\hat{\nu},\mu} \, U^{\dagger}_{l,x,\nu}.$$ Here $a$
is the step of the lattice and the lattice parameter $\beta=1/g^2$.
Physical quantities are calculated in the lattice formalism by means
of field configuration averaging, where the field configurations
(the set of all link variables) are generated with the statistical
weight $e^{-S_W}$.

 We have clarified in previous sections that additional the
Chern-Simons action describes Casimir effect. In order to find a
lattice description of Casimir interaction between boundary surfaces
let us consider Wilson loop, which describes the interaction of
charged particles. Wilson loop can be written in QED as
\begin{equation} W_C=e^{ig\oint \limits_C A_{\mu}dx^{\mu}}=e^{i\int J_{\mu}A^{\mu}dx^4}. \label{Wilsonloop} \end{equation}
The exponent in (\ref{Wilsonloop}) is the additional term to the
action. This term describes the interaction of the field $A_\mu$
with the current $J_{\mu}(x)=g\oint \limits_C
\delta(x-\xi)d\xi_{\mu}$ of charged particle. Configuration
averaging of Wilson loop $\langle W(R,T)\rangle$ (where R and T are
dimensions of the loop) converges in the limit $T\rightarrow\infty$
to:
$$\langle W(R,T)\rangle \rightarrow C e^{-V(R)T},$$ where $V(R)$ is the energy of interaction
between charged particles. The same method can be used for
calculation of Casimir energy by means of Chern-Simons action.

\begin{figure}[h]
 \begin{center} \epsfysize=60mm \epsfbox{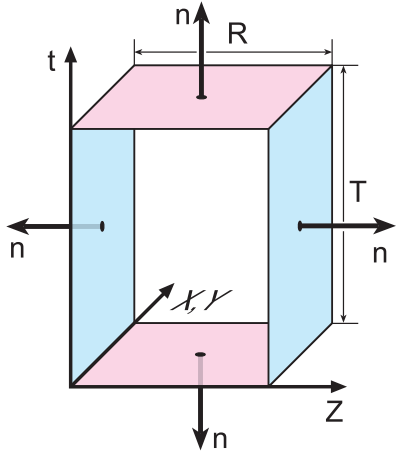}  \end{center}\caption{ Wilson bag for two plane surfaces. }
\end{figure}

Analogously to the description of charged particles interaction by
1D integral along Wilson loop, we will describe the Casimir
interaction of surfaces by corresponding 3D integral. The first
problem is that for stationary objects the action (\ref {actionCS})
is an integral from $t=-\infty$ to $t=\infty$, so by analogy to
Wilson loop, we should enclose the surface of the integration in
t-direction. The integration surface for two planes is shown in Fig.
2. This closing procedure can be performed both for plane surfaces
and for any curved surface in 3-dimensional space. As the result of
this procedure, so-called Wilson bag \cite{Seiberg,Luscher} can be
obtained. It can be written as
$$e^{i\lambda\oint\limits_{\Sigma}\varepsilon_{\mu\nu\rho\sigma}A^{\nu}F^{\rho\sigma}dS^{\mu}},$$
where $\Sigma$ is closed 3-dimensional surface in 4-dimensional
space-time. The final conclusion is that Wilson bag is (by analogy
to Wilson loop) observable quantity which gives us Casimir energy of
the objects, defined by the surface of integration. For two planes
we will calculate the following object:
\begin{equation}
 W_{Bag}(R,T)=e^{i \lambda S(R,T)},
 \label{refexp}
\end{equation}
 where $$S(R,T)= \int\limits_{0}^{T} dt
\int\!\!\int\!\!\int dx dy dz (\delta(z-R)-\delta(z))
\varepsilon_{3\nu\rho\sigma} A^{\nu}F^{\rho\sigma}+ $$
$$ +
\int\limits_{0}^{R}dz \int\!\!\int\!\!\int dx dy dt
(\delta(t-T)-\delta(t)) \varepsilon_{0\nu\rho\sigma}
A^{\nu}F^{\rho\sigma}.$$ And in the limit $T\rightarrow \infty$:
$$\langle W_{Bag}(R,T)\rangle \rightarrow C e^{-E_{cas}(R)T}.$$

The second problem is to rewrite of the Wilson bag in terms of
lattice objects (links and plaquettes). Using the expansion of link
variable:
$$U_{l,x,\mu}\approx 1+i g A_{\mu}(x) a-{1 \over 2} g^2 a^2 A_{\mu}^2+O(g^3)$$
(here $a$ is the step of the lattice) and plaquette variable:
$$U_{p,x,\mu\nu}\approx 1+i g F_{\mu\nu}(x) a^2-{1 \over 2} g^2 a^4 F_{\mu\nu}^2+O(g^3),$$
we can construct the following expression:
\begin{equation}
(1-\mbox{Re\,} (U_{p,x,\mu\nu} U_{l,x,\rho})) -(1-\mbox{Re\,}
U_{p,x,\mu\nu})-(1-\mbox{Re\,} U_{l,x,\rho}). \label{caracatica}
\end{equation}
 It can be  easily proved that expansion of (\ref{caracatica}) by
$g$ is:
\begin{equation}
g^2 a^3 F_{\mu\nu} A_{\rho} +O(g^4). \label{firstterm}
\end{equation}
 In the main order
it gives us one term in the density of Chern-Simons action
multiplied by $g^2$. After this, we replace integration by sum, and
multiple this sum by $\beta$ in order to eliminate $g^2$ in the last
expression. So we obtain a computable expression for Wilson bag.

Let us consider some results of numerical simulations. You can see
on the Fig. 3 the dependence of the Casimir energy $E_{cas}$ on $R$
(the distance between planes).
\begin{figure}[t]
\begin{center}
 \epsfbox{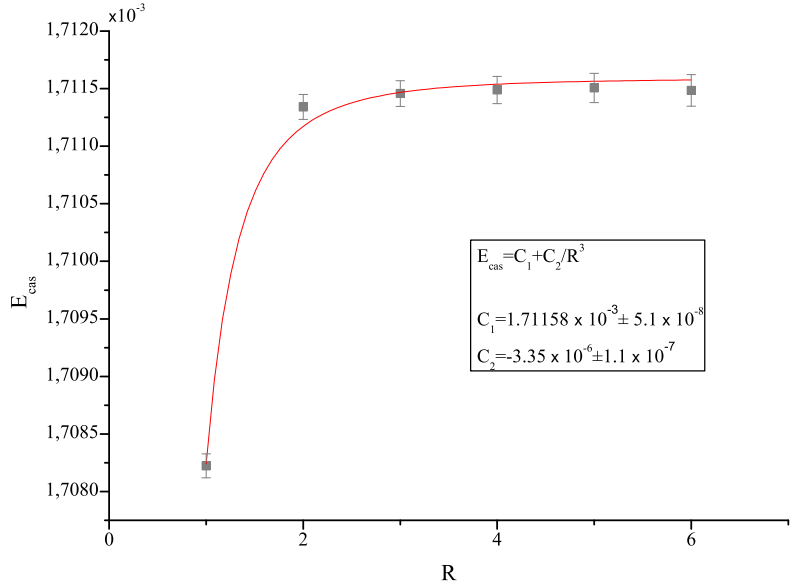} \end{center} \caption{$E_{cas}(R)$ at $\beta=8.0$ and $\lambda=0.0005$.}
\end{figure}

%\begin{figure}
%\begin{center}
% \epsfbox{beta12_1.eps} \end{center} \caption{$E_{cas}(R)$ at $\beta=12.0$}
%\end{figure}
The parameter $\beta$ is equal to 8.0 and $\lambda=0.0005$  for this
calculation. The size of the lattice is $32^4$. Unfortunately, we
can not see here $E_{cas}(R)$ dependence like $R^{-3}$. All what we
can see is a distinction of the first point from the other ones
which form the plateau. The dynamics in the plateau can not be
revealed due to numerical errors. Nevertheless, let us make a
suggestion that this difference of the first point from the plateau
is a manifestation of  the dependence of Casimir energy on the $R$
as $R^{-3}$ (this suggestion follows obviously from dimensional
reasons). So we have done a fitting of this point sequence by the
function $$E_{cas}=C_1+\frac{C_2}{R^3}.$$ The values of coefficients
$C_1$ and $C_2$ are given on the Fig. 3. Our aim is, of course, the
coefficient $C_2$. $C_1$ is an analog of something like self energy
of charged particles. Such self energy can be obtained in Wilson
loop calculation.
\begin{figure}
\begin{center}
 \epsfbox{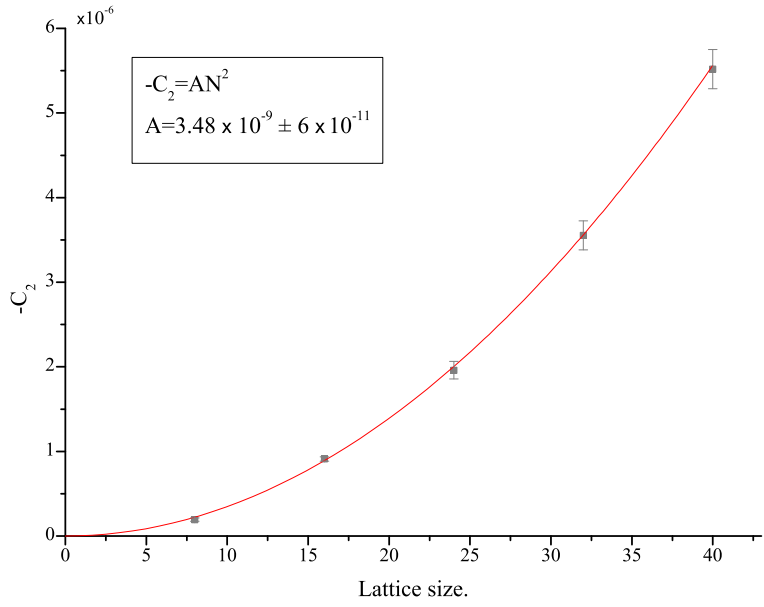} \end{center} \caption{Dependence of $C_2$ on the lattice size.}
\end{figure}
\begin{figure}
\begin{center}
 \epsfbox{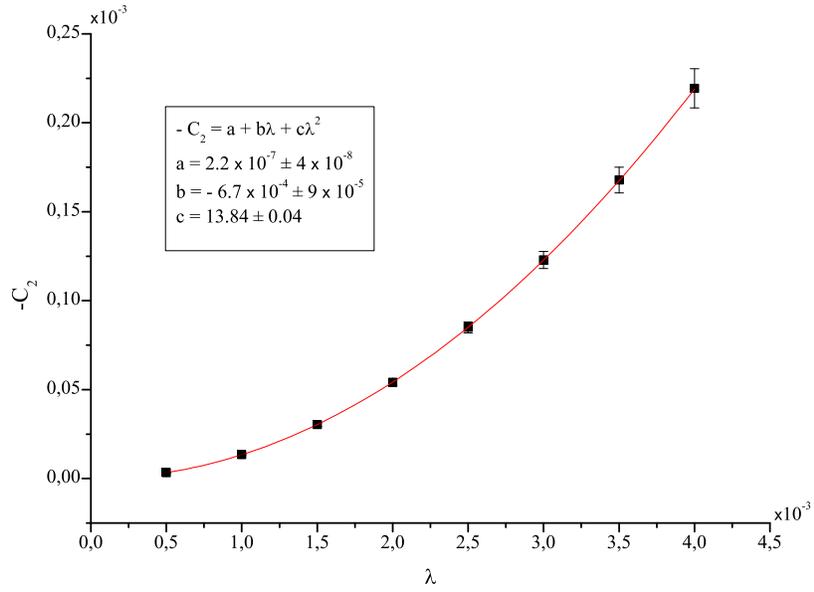} \end{center} \caption{Dependence of $C_2$ on $\lambda$.}
\end{figure}
We have performed the same calculations for different values of
$\beta$. The situation is qualitatively the same.

Now let us consider the physical meaning of the coefficient $C_2$. A
full Casimir energy of the interaction between two planes have been
obtained in these calculations. This energy is expressed in units
$a^{-1}$, where $a$ is the lattice step. If we take into account
that the area of the plane is equal to $(aN)^{2}$, where $N=32$ is
the size of the lattice, we can write the physical (dimensional)
value of the Casimir energy density:
$$E_{Cas. phys}=\frac {1}{a} \frac {C_2}{R^3} \frac {1}{(aN)^2}=
\frac {C_2 N^{-2}} {(Ra)^3}.$$
After comparison between this formula
and (\ref{fullCas}) we can conclude that ${C_2 N^{-2}}$ should be
equal to coefficient before $R^{-3}$ in (\ref{fullCas}) or in
(\ref{assimpt}) ($\lambda$ is small in our calculations by reasons
discussed later). And this coincidence is rather good. For example,
for the calculation with $\beta=8$ the expression $C_2 N^{-2} =-3.27
\times 10^{-9}\pm 1.1 \times 10^{-10}$ and the theoretical value of
this quantity for $\lambda=0.0005$ is $-3.17 \times 10^{-9}$.

The next step is the analysis of $C_2$ dependences on the size of
the lattice $N$ and on the parameter $\lambda$. The first one is
presented on the Fig. 4. We can see here that $C_2$ is proportional
to $N^2$ with high accuracy. So the calculated Casimir energy of two
planes is really proportional to theirs area.

The dependence $C_2(\lambda)$ is shown on Fig. 5. In accordance with
(\ref{assimpt}), the calculated coefficient at $R^{-3}$ is
proportional to $\lambda^2$ for small $\lambda$. Unfortunately,
present calculation methods can not give values of $C_2$ for large
$\lambda$ and we have in fact only small $\lambda$ limit. This
problem appears because the number in the exponent (\ref{refexp})
becomes too big and the errors arise for calculations with large
$\lambda$. We are studying this problem and hope that some methods
of solution can be suggested.

\begin{figure}[h]
\begin{center}
 \epsfbox{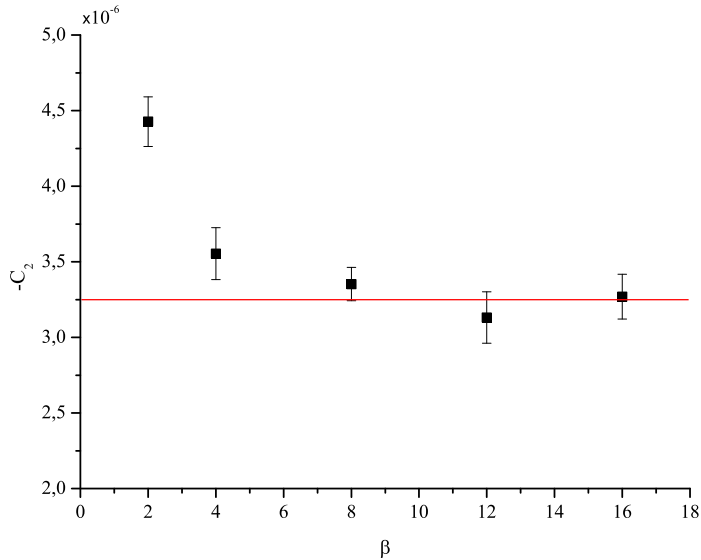} \end{center} \caption{Dependence of $C_2$ on $\beta$.}
\end{figure}

The last point is the consideration of the continuous limit of our
calculations. There are two aspects of this problem. First, when we
calculate Wilson bag, in fact, only the special quantity being
approximately equal to Chern-Simons action is calculated
(\ref{caracatica}).
 Expansions of all quantities have been done by powers of $g$. So the smaller is $g$, the closer is our
 lattice Wilson bag quantity (\ref{caracatica}) to its first term in expansion (\ref{firstterm}).
Thereby calculations in the limit $g\rightarrow 0$ or $\beta
\rightarrow \infty$ are the first stage in continuous limit. This
stage is shown in Fig. 6 for $C_2$ and in Fig. 7 for $C_1$.
Additional line at Fig. 6 is the theoretical value of $C_2$ for
$\lambda=0.0005$. As we can see in the limit $\beta \rightarrow
\infty$ the coefficient $C_2$ becomes stable around the right
analytical value. This stability can be considered as an additional
argument for a suggestion that $C_2$ has the real physical meaning.
As for $C_1$, it hasn't any continuous limit due to permanent growth
with increasing of $\beta$.

\begin{figure}[t]
\begin{center}
 \epsfbox{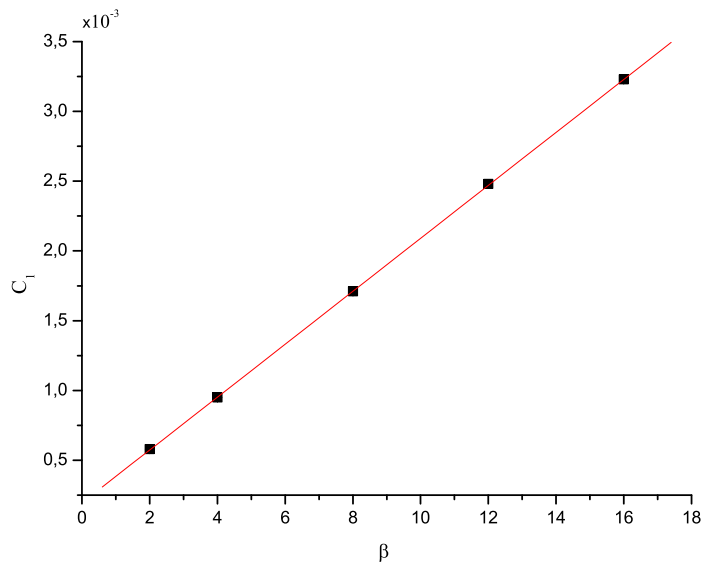} \end{center} \caption{Dependence of $C_1$ on $\beta$.}
\end{figure}

The second stage is "infinite lattice volume" limit $N \rightarrow
\infty$. But our method is not very sensitive to the lattice size
for the plane form of the boundaries. Even for rather small lattices
(see Fig. 4) the value of the coefficient before $R^{-3}$ is close
to the right quantity. There is one peculiarity in this continuous
limit procedure and we think that this moment should be emphasized.
The lattice step $a$ disappears from our considerations because the
final result of our calculation is dimensionless coefficient at
$R^{-3}$ in expression for Casimir energy per unit area. So the
value of $a$ is not important for us and instead of limit $a
\rightarrow 0$ we have limit $N \rightarrow \infty$.

%%%%%%%%%%%%%%%%%%%%%%%%%%%%%%%%%%%%%%%%%%%%%%%%%%%%%%%%%%%%%%%%%%%%%%%%%%%%%%%
\section*{Conclusions}

In this paper we have proposed the numerical method for the Casimir
energy calculation based on the lattice simulations of QED. We have
combined two ideas: the generation of the boundary conditions by
means of the additional Chern-Simons boundary action and the lattice
"Wilson bag" concept (the lattice presentation of the closed
3-dimensions surface in Euclidian 4-dimensions space). This
combination is in fact a lattice definition of the respective
quantum observable for the Casimir energy and for the Casimir
interaction between surfaces. We have tested our method in the
simplest case of the Casimir interaction between two plane surfaces
and have achieved a good agreement with the analytical results for
this problem.

A big advantage of our method is its universality. This method can
be applied for following calculations:
\begin{itemize}
\item Investigation of the Casimir interaction between surfaces of
complicated shapes which are interesting from the experimental point
of view.

\item Calculation of thermal corrections to the Casimir force.
(Unfortunately, this very important problem is very difficult for
any analytical investigations in physically interesting cases.)

\item Study of Casimir effects for non-abelian gauge fields and
for fermions which is very essential for phenomenological models in
heavy-ion collisions and for the models of low-energy hadron states
\cite{Hosaka, sveshnikov}.
\end{itemize}
These problems are under consideration now.

%%%%%%%%%%%%%%%%%%%%%%%%%%%%%%%%%%%%%%%%%%%%%%%%%%%%%%%%%%%%%%%%%%%%%%%%%%%%%%%
\section*{Acknowledgments}

We thank  Prof. V.V. Nesterenko, Dr. I.G. Pirozhenko and Dr. V.N.
Marachevsky for interesting discussions about the Casimir problem
and Andrew Zayakin for essential remarks about vortex dynamics of
the Chern-Simons theory. The resources of MSU supercomputer center
were used in our calculations. The work is partially supported by
the Russian Federation President's Grant 195-2008-2.

\end{document}